\documentclass[fleqn,usenatbib]{mnras}

\usepackage{newtxtext,newtxmath}
\usepackage[T1]{fontenc}

\DeclareRobustCommand{\VAN}[3]{#2}
\let\VANthebibliography\thebibliography
\def\thebibliography{\DeclareRobustCommand{\VAN}[3]{##3}\VANthebibliography}

\usepackage{graphics}
\usepackage{tabu}
\usepackage{natbib}
\usepackage{color}
\usepackage{rotating}
\usepackage{multirow}
\usepackage{footnote}
\usepackage{datetime}
\usepackage{amsmath}
\usepackage{mathrsfs}
\usepackage[normalem]{ulem}

\usepackage{lipsum} 
\title[Jellyfish Tail Turbulence]{Turbulence in the Tail of a Jellyfish Galaxy}

\author[Y. Li et al.]{
Yuan Li,$^{1}$\thanks{E-mail: yuan.li@unt.edu}
Rongxin Luo,$^{2}$
Matteo Fossati,$^{3,4}$
Ming Sun$^{2}$
and Pavel J\'achym$^{5}$
\\
$^{1}$Department of Physics, University of North Texas, Denton, TX 76203, USA\\
$^{2}$Department of Physics and Astronomy, University of Alabama in Huntsville, Huntsville, AL 35899, USA\\
$^{3}$Dipartimento di Fisica G. Occhialini, Universit\`a degli Studi di Milano-Bicocca, Piazza della Scienza 3, 20126 Milano, Italy\\
$^{4}$INAF-Osservatorio Astronomico di Brera, via Brera 28, I-20121 Milano, Italy\\
$^{5}$Astronomical Institute of the Czech Academy of Sciences, Bo\v{c}n\'i II 1401, 14100, Prague, Czech Republic\\
}

\date{Accepted XXX. Received YYY; in original form ZZZ}

\pubyear{2022}

\begin{document}
\label{firstpage}
\pagerange{\pageref{firstpage}--\pageref{lastpage}}
\maketitle

\begin{abstract}
When galaxies move through the intracluster medium (ICM) inside galaxy clusters, the ram pressure of the ICM can strip the gas from galaxies. The stripped gas forms tails on the trailing side. These galaxies are hence dubbed ``jellyfish galaxies''. ESO 137-001 is a quintessential jellyfish galaxy located in the nearest rich cluster, the Norma cluster. Its spectacular multiphase tail has complex morphology and kinematics both from the imprinted galaxy's interstellar medium (ISM) and as a result of the interactions between the stripped gas and the surrounding hot plasma, mediated by radiative cooling and magnetic fields. We study the kinematics of the multiphase tail using high-resolution observations of the ionized and the molecular gas in the entire structure. We calculate the velocity structure functions (VSFs) in moving frames along the tail and find that turbulence driven by Kelvin-Helmholtz (KH) instability quickly overwhelms the original ISM turbulence and saturates at $\sim 30$ kpc. There is also a hint that the far end of the tail has possibly started to inherit pre-existing large-scale ICM turbulence likely caused by structure formation. Turbulence measured by the molecular gas is generally consistent with that measured by the ionized gas in the tail but has a slightly lower amplitude. Most of the measured turbulence is below the mean free path of the hot ICM ($\sim 11$ kpc). Using warm/cool gas as a tracer of the hot ICM, we find that the isotropic viscosity of the hot plasma must be suppressed below $0.01\%$ Spitzer level.

\end{abstract}


%%%%%%%%%%%%%%%%%%%%%%%%%%%%%%%%%%%%%%%%%%%%%%%%%%

%%%%%%%%%%%%%%%%% BODY OF PAPER %%%%%%%%%%%%%%%%%%
\section{Introduction}
\label{sec:intro} 
\setcounter{footnote}{0} 

Jellyfish galaxies are galaxies with ram pressure-stripped tails typically found in galaxy clusters. When galaxies fall into galaxy clusters, the ram pressure of the ICM strips the galaxy's interstellar medium (ISM) and circum-galactic medium (CGM) to form a trailing tail \citep[see][and references therein]{2022A&ARv..30....3B}. While the galaxy itself experiences dramatic transformation from late-type to early-type as a result of ram pressure stripping of cool ISM and subsequent quenching of star formation, the gas in the tail interacts with the surrounding ICM, forming a turbulent multiphase structure. The rich physical processes (star formation, turbulence, magnetic fields, etc) involved in these interactions make jellyfish tails unique labs to study turbulent multiphase gas and plasma.

ESO 137-001 is one of the nearest jellyfish galaxies with the richest amount of multi-wavelength data. It is located in the Norma cluster (Abell 3627), the closest rich cluster, with a projected distance of $\sim 200$ kpc from the cluster center \citep{Sun2006, Pavel2014}. ESO 137-001 has a very small line-of-sight velocity difference ($\sim 200$ km/s) from the cluster mean, suggesting that its motion is mainly in the plane of the sky. The tail of ESO 137-001 extends beyond $\sim 80$ kpc. Its multiphase components have been observed in multiple wavelengths. Its hot X-ray gas has been observed with Chandra, and its molecular component has been observed with Spitzer \citep{2010ApJ...717..147S} and the Atacama Large Millimeter/submillimeter Array (ALMA) CO \citep{Pavel2019}. 
The ionized gas in the tail has been observed with the Multi-Unit Spectroscopic Explorer (MUSE) over the years, showing that the multiphase structure also has complex kinematics \citep{Fossati2014, Fossati2016, Luo2022}. %Hubble

Ram pressure stripping of disk galaxies has been studied extensively using numerical simulations both in cosmological context \citep{Yun2019, Troncoso-Iribarren2020} and in idealized setups, many of which have been modeled after ESO 137-001 \citep{Roediger2006, Tonnesen2012, Ruszkowski2014, 2022MNRAS.512.5927F}. On smaller scales, many numerical studies have been conducted to better understand the detailed interactions between moving cool clouds and their surrounding hot gas \citep[e.g.,][]{2018MNRAS.480L.111G, 2019MNRAS.487..737J, 2022ApJ...925..199A, Gronke2022}, which are basic building blocks of jellyfish tails that cannot be studied with sufficient resolution in global simulations. These focused small-scale numerical models have revealed a complex interplay between radiative cooling and turbulence driven by Kelvin-Helmholtz (KH) instability that can be further complicated by the inclusion of magnetic fields \citep{Berlok2019, 2020MNRAS.492.1841L, Mandelker2020, 2020ApJ...892...59C}. 

In this work, we analyze the velocity structure function (VSF) of the multiphase tail of ESO 137-001. 
We describe the data used in this work and how we compute VSFs in Section~\ref{sec:data}. In Section~\ref{sec:results}, we analyze the origin of the turbulent motion in the tail (Section~\ref{sec:frames}), compare the kinematics of the ionized gas and the molecular gas (Section~\ref{sec:CO}), discuss possible biases and uncertainties (Section~\ref{sec:uncertainties}), and use the measured turbulence to constrain the isotropic viscosity of the hot ICM (Section~\ref{sec:viscosity}). We conclude our work in Section~\ref{sec:conclusions}. 

\section{Data Processing and Analysis Method}\label{sec:data}

The ionized gas of ESO 137-001 was observed using the MUSE wide-field mode with a spatial sampling of $0.2''$ and a spectral resolution of $\rm R\sim 2600$. The data used in this work was collected from multiple observational programs spanning over several years. The seeing ranges from $0.57''$ to $1.94''$ with a median of $1.04''$. A Gaussian profile was adopted to fit each line of the $\rm H\alpha+[NII]$ complex. The velocity and velocity dispersion of [NII] lines are tied to that of the Ha line ( i.e. assuming the three lines have common kinematic parameters). A median filter with a kernel of $4\times4$ spaxels ($0.8'' \times 0.8''$) was also applied to smooth the data cube.
We discuss the effects of the point-spread-function (PSF) in Section~\ref{sec:uncertainties}. Since we focus on the diffuse gas in the tail, HII regions and the regions contaminated with foreground stars are also masked. More details on MUSE data reduction can be found in \citet{Fossati2014, Fossati2016, Luo2022}. 

\begin{figure}
\centering
\includegraphics[scale=0.31,trim=0.5cm 0cm 0cm 0cm, clip=true]{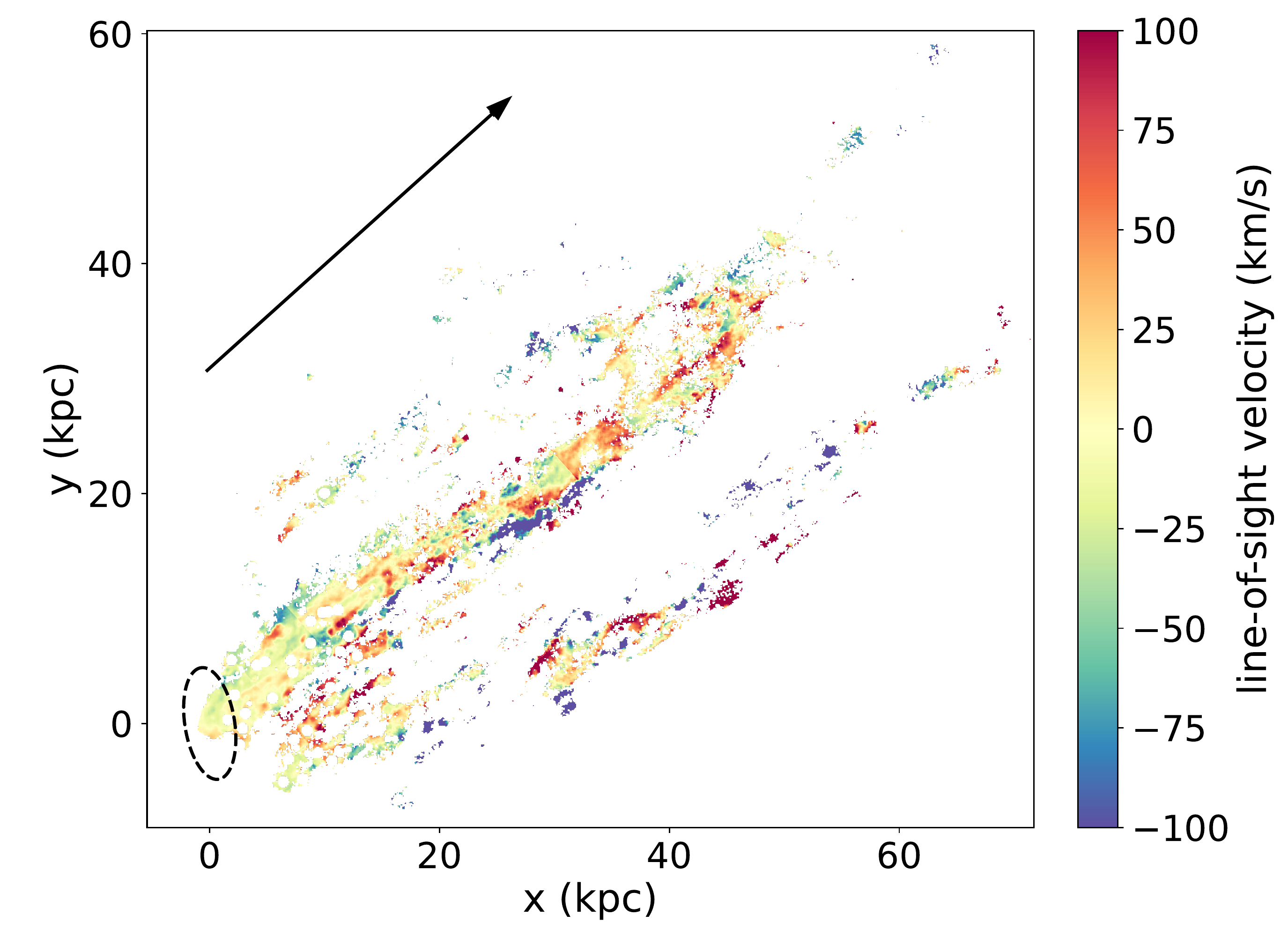}
\caption{Residual velocity map of the H$\alpha$ clouds in ESO 137-001 observed with MUSE. The motion of the ionized gas appears turbulent. The map is obtained by subtracting rotation from the original line-of-sight velocity map (Section~\ref{sec:data}). More details of rotation modeling can be found in \citet{Luo2022}. The arrow with a position angle of $\sim -48^{\circ}$ is the direction of the tail according to the kinematic modeling, and the length of the arrow represents the size of the moving frame used in our analysis (Section~\ref{sec:frames}). The MUSE PSF is slightly smaller than the size of the arrow head. The ellipse presents the half-light radius of ESO 137-001 from the HST F160W image.
}\label{fig:map}
\end{figure}

As is shown in \citet{Fossati2014}, the ionized gas near the head of the tail presents a clear velocity gradient roughly perpendicular to the stripping direction, which is considered as the imprint of the galactic rotation. To remove this rotation imprint, \citet{Luo2022} divide the velocity field into several regions along the stripped tails and model the velocity gradient in each region separately. The residual velocity field, shown in Figure~\ref{fig:map}, is obtained by subtracting these modeled gradients from the observed velocity field. \citet{Luo2022} used two methods to remove rotation and suggested that using local velocity gradients (adopted here) is a better method than using the global velocity gradient. We have verified that our main conclusions are not affected by the method of rotation modeling.

Most of the ionized gas have small uncertainties ($\sim 10\, \rm km\, s^{-1}$) associated with their measured line-of-sight velocities. To reduce noise in our analysis, we first apply a mask that removes all the data with signal-to-noise ratio (SNR) less than 3 for the H$\alpha$ line. We then remove all pixels with velocity errors larger than 22 km/s, which comprise $\sim 15$\% of the remaining data. Most of these pixels are located at the edge of clouds. Our results are not sensitive to the exact choice of the velocity error threshold (see Appendix for more detailed discussions). 

The molecular gas in ESO 137-001 was observed with ALMA traced by CO(2-1) emission. The observations were conducted with a velocity resolution of $\sim 0.64\, \rm km\, s^{-1}$ and the final data cube has a pixel size of $0.14''$, slightly smaller than that of the MUSE data. The synthesized angular resolution is $\sim 1.4'' \times 1.2''$, similar to the PSF of MUSE. Details about the processing of ALMA data are described in \citet{Pavel2019}. We exclude pixels with flux below 0.05 $\rm Jy \, beam^{-1} \, km \, s^{-1}$ to reduce noise in our analysis.

We compute the first-order VSF of both the ionized gas and the molecular gas in ESO 137-001 to study the nature of their motion. The VSF is computed as follows: for each pair of pixels, we record their spatial separation $\ell$ in the projected plane of the sky and their velocity difference $\delta v$. We then compute the average absolute value of the velocity differences $\langle |\delta v| \rangle$ within bins of $\ell$ to obtain the VSF. VSF is related to the turbulent kinetic energy power spectrum. For Kolmogorov turbulence, the first-order VSF has a slope of $1/3$ within the inertial range, and for supersonic turbulence, the slope is $1/2$.

\section{Results and Discussions}\label{sec:results}

\subsection{The Development of Turbulence in the Tail}\label{sec:frames}

As ESO 137-001 falls into the galaxy cluster, ram pressure of the ICM strips the gas in the ISM and the CGM out of ESO 137-001 to form a jellyfish tail. In addition to the bulk proper motion of the galaxy, the stripped gas originally also has a rotation and turbulent motion in the galaxy's rest frame. After the gas is stripped from the galaxy, its velocity is not decelerated immediately. The shear between the stripped gas and the surrounding ICM can cause KH instability to develop. If the cool gas becomes kinematically coupled to the hot ICM, it can also pick up the pre-existing ICM turbulence. 

To understand how turbulence develops in the tail of ESO 137-001, we use a 30 kpc-wide rectangular moving frame placed along the tail direction. We center the first frame at $\sim 10$ kpc from the galaxy in the upstream direction (the opposite side of the tail), allowing it to only include the ``head'' of the whole structure, and move the frame along the tail vector (arrow in Figure~\ref{fig:map}) by 5 kpc with every step. At each step, we compute the VSF of all the pixels within the frame. The width of the moving frame is chosen to be large enough to cover a wide dynamical range in $\ell$, but still small enough so that the differences between different frames are easy to see. 
The width of the frame is similar to the width of the whole structure so that the pixels within most frames are reasonably evenly distributed within a square. We have experimented with different frame sizes. A wider frame would reduce the differences between different VSFs while a narrower frame shows the differences more dramatically. However, a narrower frame limits the reliably probed dynamical range to the width of the frame. We thus chose a frame size that shows a clear evolution in the VSFs but also allows us to probe the VSFs at large $\ell$. The shape of the VSFs and the trend of the evolution are not sensitive to the exact frame size.

\begin{figure}
\centering
\includegraphics[scale=0.35,trim=0.5cm 0cm 0cm 0cm, clip=true]{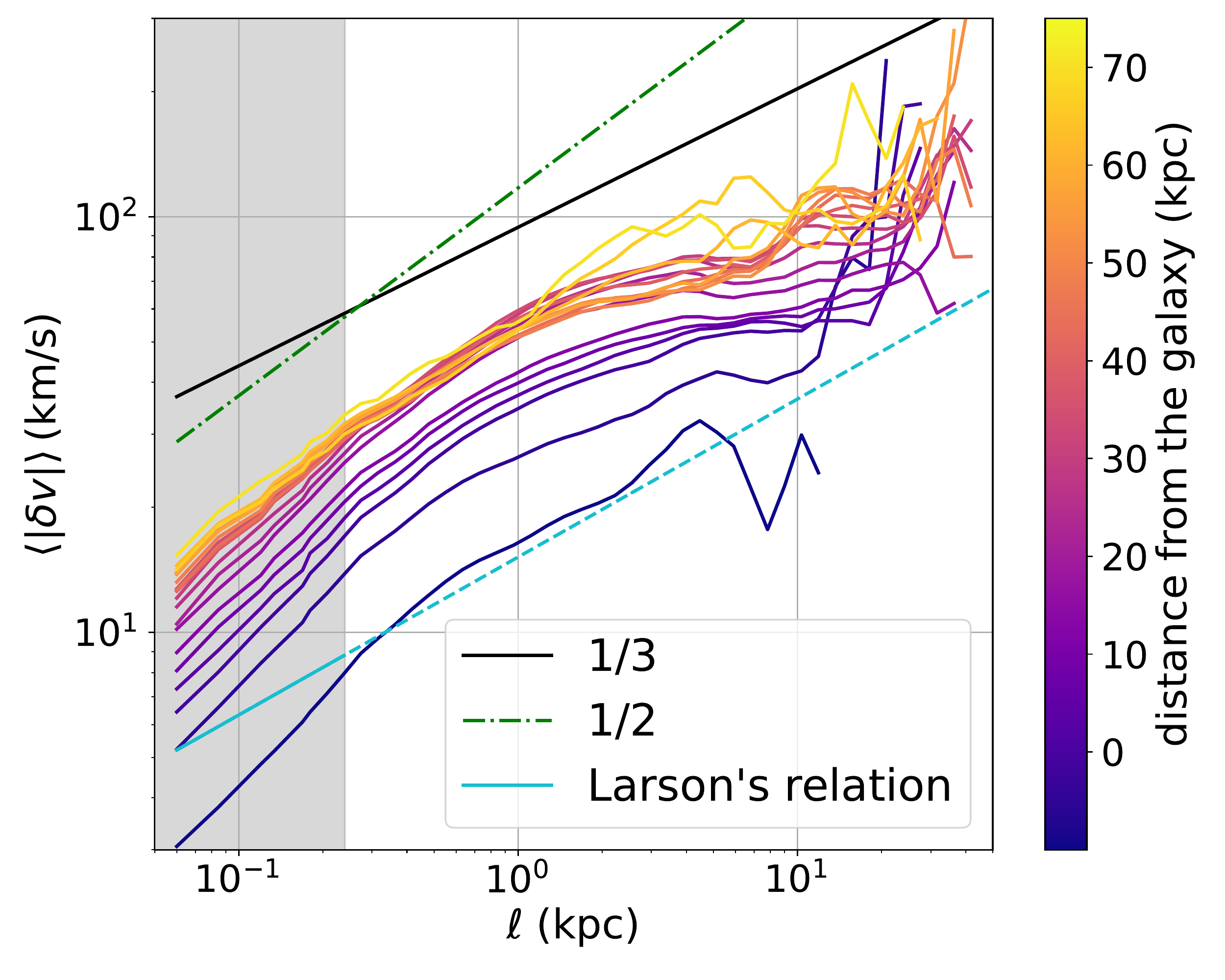}
\caption{The buildup of turbulence in the tail of ESO 137-001. First-order VSFs are computed within 30 kpc-wide moving frames along the tail direction, color-coded by the distance from the galaxy to the center of the frame. The first frame is placed in the upstream direction such that it only covers the head of the structure (mainly the disk galaxy itself). The shaded region denotes where PSF can steepen the VSF. }\label{fig:frames}
\end{figure}

The evolution of the VSF is shown in Figure~\ref{fig:frames}. In all parts of the tail, the VSF shows a similar overall shape: on small scales ($\ell \lesssim 1-2 $ kpc), the VSF follows a power law that spans about an order of magnitude in $\ell$, consistent with the expectation of a turbulent flow. The slope is close to 1/3 (Kolmogorov turbulence) within the inertial range that is reliably measured (above the shaded region). However, the exact slope can be affected by observational effects discussed in Section~\ref{sec:uncertainties} and in the Appendix. The intrinsic slope of ICM turbulence may also deviate from Kolmogorov as a result of plasma instabilities \citep[e.g.,][]{2022arXiv220705189A}. The shaded region indicates where the VSF can be steepened by the PSF (see more discussion in Section~\ref{sec:uncertainties}). On larger scales  ($\ell \gtrsim 1-2 $ kpc), the VSF flattens, which we interpret as the energy injection scales. 

Figure~\ref{fig:frames} shows that the level of turbulence gradually builds up from the head to the tail and appears to saturate at $\sim 30$ kpc. Since the first frame only covers the head, the VSF mainly shows the turbulence of the ISM of the late-type galaxy itself. Its shape and amplitude are very similar to Larson's relation (extrapolated to larger scales) found for molecular clouds in the Milky Way \citep{Larson1981}. The increase in the level of turbulence (the amplitude of the VSF) means that turbulence in the tail of ESO 137-001 is not dominated by the ``frozen-in'' ISM turbulence. Instead, there are additional physical processes that enhance the turbulent motion of the stripped gas. 

We first consider the development of KH instability. The gas in the H$\alpha$ tails first originates from the ISM and the CGM of the galaxy. As it interacts with the ICM, it is decelerated (or accelerated by the hot ICM wind in the galaxy's rest frame) via mass and momentum exchanges. The shear between the streams of cool gas and the surrounding hot ICM creates KH instability, which generates turbulence. Recent simulations of idealized radiative cold streams find that KH instability saturates at $\sim 20 L/c_s$, where $L$ is the diameter of the stream and $c_s$ is the sound speed of the hot surrounding medium \citep{Berlok2019, Mandelker2020}. At the location of the tail, the hot ICM has $c_s \sim 10^3$ km/s \citep{Sun2010}. The width of individual H$\alpha$ tails is $3-4$ kpc \citep{Sun2007}. This gives us an estimated KH saturation time to be around $60-80$ Myr. If the saturation of turbulence at $\sim 30$ kpc is due to the saturation of KH instability, then from 0 to 30 kpc, the average velocity of the gas along the tail direction is about $400-500$ km/s in the galaxy's rest frame. This is in good agreement with what is found in numerical simulations modeled after ESO 137-001 \citep{Tonnesen2012, Tonnesen2014}. The amplitude of turbulence is also generally consistent with the theoretical expectations of saturated KH instability. \citet{Berlok2019} show that in a magnetized medium, the kinetic energy perpendicular to the stream in the saturation phase is on the order of $\lesssim 1\%$ of the initial parallel kinetic energy. The velocity of ESO 137-001 is estimated to be $\sim 1000-2000$ km/s \citep{Sun2006, Fossati2014}, although the models in \citet{Pavel2014} suggest that the orbital velocity of ESO 137-001 has to be higher than 3000 km/s. This implies a turbulent velocity of $\lesssim 100-300$ km/s, consistent with our measurements. Thus the development of KH instability is a plausible explanation for the evolution of the VSF we see in Figure~\ref{fig:frames}. 

We now consider another possible source of turbulence -- the large-scale ICM turbulence. Structure formation and accretion can drive turbulence in the bulk of the ICM, as is shown in numerical simulations \citep[e.g.,][]{Ryu2008, Vazza2009, ZuHone2013, Shi2018, Angelinelli2020}. As the stripped gas interacts with the ICM, both mass and momentum can be exchanged. The gas in the tail may pick up pre-existing ICM turbulence from the accretion of gas and merging substructures. The driving scales for these processes are typically at hundreds of kpc or even larger \citep[e.g.,][]{Dolag2005}. Figure~\ref{fig:frames} shows a flattening of the VSF around $1-10$ kpc, suggesting an energy injection at these scales. Since the widths of the tails are typically a few kpc, we have also considered the possibility that the flattening is related to the width limit. We compute the VSFs of pairs of pixels along the tail direction (not limited by the width) and perpendicular to the tail direction (width limited). They have similar shapes and both flatten around $1-10$ kpc. Thus below $1-10$ kpc, the turbulence of the tail is unlikely to be dominated by the cascade of large-scale ICM turbulence. 

At $\sim 10-20$ kpc, the VSF of the distant part of the tail ($>40$ kpc from the galaxy, corresponding to the red/orange lines in Figure~\ref{fig:frames}) bends upward, indicating an additional source of energy injection. The amplitude also continues to grow toward the end of the tail, instead of saturating at $\sim 30$ kpc as the VSFs on smaller scales do. Therefore, the behavior of the VSFs at large $\ell$ likely reflects the pre-existing large-scale ICM turbulence. We caution though that the sampling size at large $\ell$ becomes rather limited, and thus the VSF is less robustly measured. 

Future studies of more jellyfish tails, as well as the newly discovered orphan cloud \citep{Ge2021} may help further disentangle the roles of KH instability and large-scale ICM turbulence in driving the motions of the stripped gas. Analyzing numerical simulations of jellyfish galaxies can also be helpful, especially by comparing idealized and cosmological simulations, as the former does not have any large-scale ICM turbulence. 

\subsection{Turbulence Traced with Molecular Gas}\label{sec:CO}

\begin{figure}
\centering
\includegraphics[scale=0.33,trim=0.5cm 0cm 0cm 0cm, clip=true]{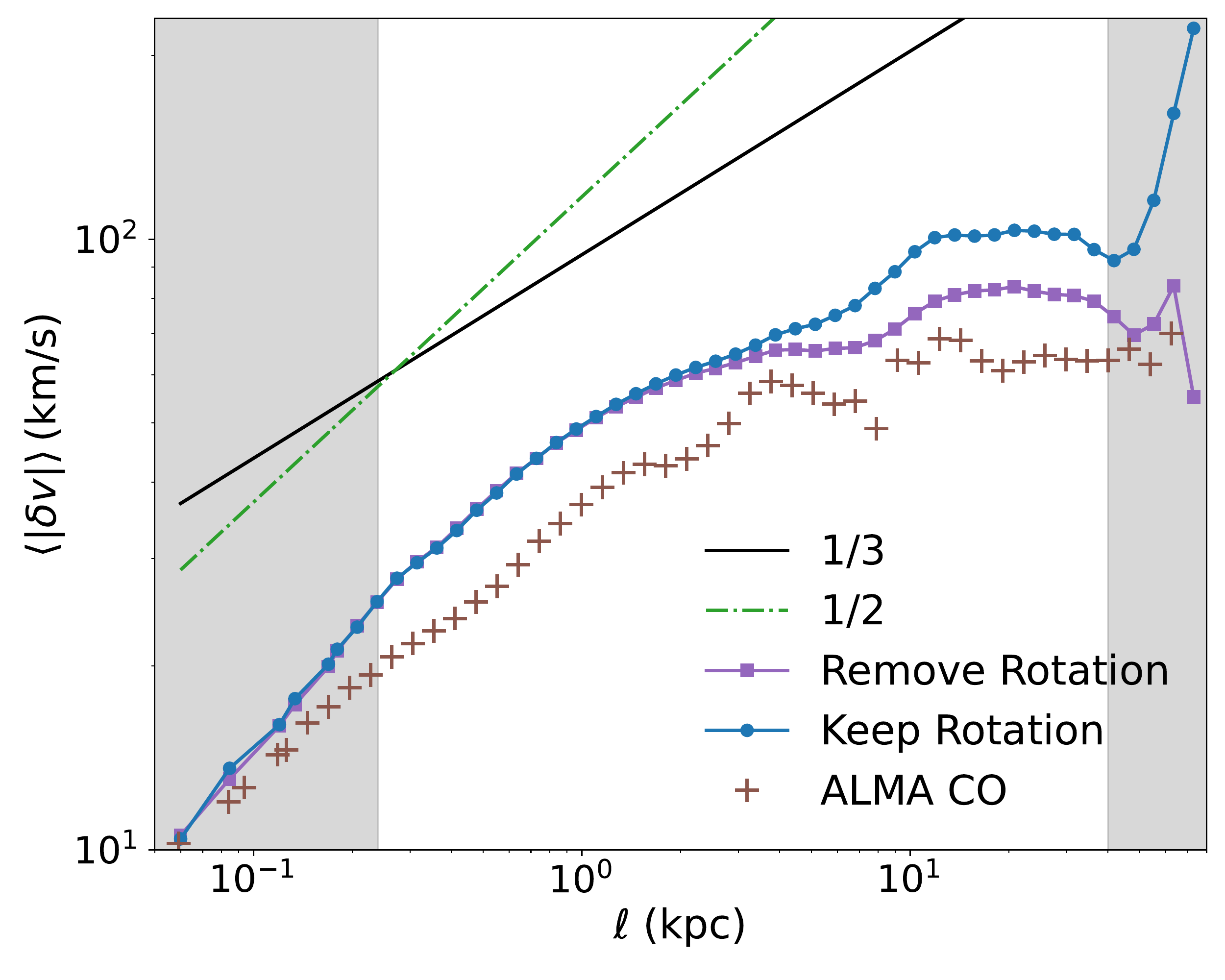}
\caption{The VSFs of the entire H$\alpha$ structure with rotation removed (purple) and for the original data with rotation (blue). Also shown is the VSF of the molecular gas observed with ALMA. The shaded region at small $\ell$ denotes where PSF can steepen the VSF. The shaded region at large $\ell$ is where sampling becomes very limited. 
}\label{fig:CO}
\end{figure}

The cold molecular gas in ESO 137-001, traced by CO(2-1) emission, has been observed with ALMA \citep{Pavel2019}. The overall spatial extension of the molecular gas is similar to that of the ionized gas, but the covering fraction of CO is much smaller, as CO emission comes only from the densest cores of the multiphase structure. The SNR of the ALMA data is also relatively low compared with the MUSE data. Thus we do not attempt to model and remove rotation for the molecular gas. Figure~\ref{fig:CO} shows that for the H$\alpha$ gas, the VSFs with and without rotation only mildly differ on scales larger than $\sim 3-4$ kpc, where the original data with rotation shows a higher amplitude in its VSF, as one would expect. 

Overall, the VSF of the molecular gas traces that of the ionized gas, suggesting a reasonably good kinematic coupling between the two phases. The amplitude of the CO VSF is lower, even compared with the VSF of the ionized gas without rotation. This may be partly related to a sampling bias. There is very little CO emission detected with high S/N far from the galaxy. As Figure~\ref{fig:frames} shows, the VSF of the gas close to the galaxy has a lower amplitude. It may also indicate an imperfect kinematic coupling between the cold dense molecular gas and the ionized gas. For example, \citet{Gronke2022} show that in idealized simulations of a multiphase turbulent medium, the VSF of the cold phase has a lower amplitude than that of the hot phase. A similar trend is also found by \citet{Mohapatra2022} in both hydrodynamic and magnetohydrodynamic simulations. Analyses of the Milky Way molecular complexes show that the VSF of the H$\alpha$ gas can be higher than that of the cold molecular gas when fresh energy is injected preferentially into the warm ionized phase via stellar feedback \citep{Ha2022}. 

We emphasize again that the CO data is relatively noisy. The exact amplitude and shape of the VSF are somewhat sensitive to the data cleaning process. 
The sparse spatial coverage of CO also limits the sampling statistics, which likely causes some of the ``bumps'' in the VSF. The general trend of the result is reliable, but we caution against over-interpreting the detailed features in the CO VSF.

\subsection{Limitations and Uncertainties}\label{sec:uncertainties}

In this section, we discuss three main sources of limitations and uncertainties in our analysis: the PSF and two projection effects. 

The VSF of the ionized gas can be affected by the PSF. \citet{Li2020} conduct a ``double-seeing'' experiment where they smooth the Virgo MUSE data using a Gaussian kernel with a full width at half maximum (FWHM) equal to the observed one. The resultant VSF shows only a mild suppression compared with the original VSF. \citet{Chen2022} carry out an expansive analysis and show that for VSFs with shallower intrinsic slopes (Virgo VSF has a very steep slope of $\sim 1$), the effects of PSF smoothing can be larger. The VSF can bend downward significantly near the FWHM for fainter objects such as jellyfish tails. 
The H$\alpha$ VSF in our analysis indeed steepens near the FWHM (shaded region in the figures).

The effects of projection can be rather complicated. When the emission comes from ``point sources'', the 2D projected VSF of a 3D structure has a shallower slope compared with the intrinsic one. This is because two points close to each other in projection may be well separated along the line-of-sight, and thus have a large velocity difference \citep{Xu2020FRB, Zhang2022}. The thicker the 3D structure is along the line-of-sight, the stronger this projection effect becomes \citep{Qian2015}. Another projection effect comes from multiple emitting clouds along a single line-of-sight. If the velocity is obtained by a one-component fit or by taking a flux-weighted mean, the resulting VSF is steeper than the intrinsic one \citep{Chen2022}. This is because taking an average is effectively smoothing the velocity field. The more overlapping clouds there are along individual lines of sight, the stronger this projection effect is. The extreme case is volume-filling gas, such as the X-ray emitting plasma, the projection bias of which has been studied both analytically and numerically \citep{ZuHone2016, Xu2020}. 

In the case of ESO 137-001, the second projection effect (steepening VSF) is likely very small. The clouds in the tail are reasonably sparse and not volume filling. We only noticed $\sim 6\%$ of the pixels with multiple line-of-sight components. Furthermore, although we use one-component fitting, the fit is usually only sensitive to the strongest component. \citet{Chen2022} show that the VSF steepening as a result of this projection effect is subtle even for quasar host nebulae where the emitting clouds are more volume-filling than jellyfish tails. The first projection effect (flattening VSF) is harder to assess properly without knowing the 3D structure of the tails. 
The molecular gas has a smaller spatial coverage than the ionized gas, but the two components have similar slopes in their reliably measured inertial ranges. 
This suggests that the projection bias does not significantly affect the slopes of the VSFs in our analysis. Studying simulated jellyfish tails can further help us understand and potentially correct for the two projection effects in the future.

\subsection{Suppressed Viscosity in the Intracluster Plasma}\label{sec:viscosity}

The ICM is a weakly magnetized and weakly collisional plasma, where the Coulomb electron mean free path, $\lambda_{mfp}$, is comparable to the scales of interest. For example, in cool-core cluster centers, $\lambda_{mfp}$ is on the order of hundreds of pc, and in cluster outskirts, $\lambda_{mfp}$ is as large as tens of kpc. 
Plasma instabilities happen on much smaller scales (e.g., the Larmor radii of electrons and ions are below npc scales) that are impossible to directly observe. Numerical simulations of kinetic plasma processes cannot reliably predict the plasma behavior on large scales due to the huge dynamical range. 
Nonetheless, these microscopic plasma instabilities may affect the effective transport coefficients (e.g., conduction and viscosity) and therefore impact the physical properties of the ICM on large scales \citep{Kunz2014} that can be probed with observations. 

The effective viscosity of the ICM has been probed previously with Chandra X-ray observations of the hot plasma. For example, X-ray surface brightness fluctuations have been used to infer the power spectrum of turbulence in the ICM. If the scales probed are close to the turbulent dissipation scale (due to Spitzer viscosity), then the existence of turbulence can put constraints on the effective viscosity \citep{Zhuravleva2019}. ``Cold fronts'', which are contact discontinuities between cooler and hotter plasmas as a result of merging substructures, have also been used to study ICM microscopic properties by examining the KH instability at the interfaces \citep{ZuHone2016}. KH instability can also develop in the wake of galaxies falling into galaxy clusters. If the in-falling galaxy is an early-type galaxy, its own CGM should be detectable in the X-ray but has a lower temperature than the ICM. The interaction between the ICM and the CGM generates KH rolls that can be observed with Chandra. The size of these KH rolls, as well as the length of the whole tail, have been used to constrain the effective ICM viscosity to be below $\sim 5-10\%$ of the Spitzer level \citep{2015ApJ...806..104R, 2017ApJ...834...74S}. 

Observations of multiphase gas in cluster centers suggest that different temperature components are well-coupled within the dynamical range we can probe \citep{Gendron-Marsolais2018, Li2020}. Numerical simulations of turbulent multiphase gas also show that the hot and cool components are dynamically coupled, although their VSFs can differ in the exact amplitudes and slopes \citep{Wang2021,Mohapatra2022,Gronke2022}. \citet{Li2020} use H$\alpha$ filaments in cluster centers as tracers of the X-ray ICM, and show that isotropic viscosity of the hot plasma must be suppressed. Because of the relatively small $\lambda_{mfp}$ in cluster centers, the constraint on viscosity is at a level similar to the previous constraint based on X-ray surface brightness fluctuation analysis of the Coma cluster. Outside of cluster cool cores, the temperature is higher and the density is lower. Thus $\lambda_{mfp}$ can be much larger. The Kolmogorov microscale, where the turbulent kinetic energy is dissipated into heat due to isotropic viscosity, is also larger. At the location of ESO 137-001, $\lambda_{mfp}\sim 11$ kpc, which allows us to put a much better constraint on the level of isotropic viscosity of the ICM.

\begin{figure}
\centering
\includegraphics[scale=0.37,trim=0.5cm 0cm 0cm 0cm, clip=true]{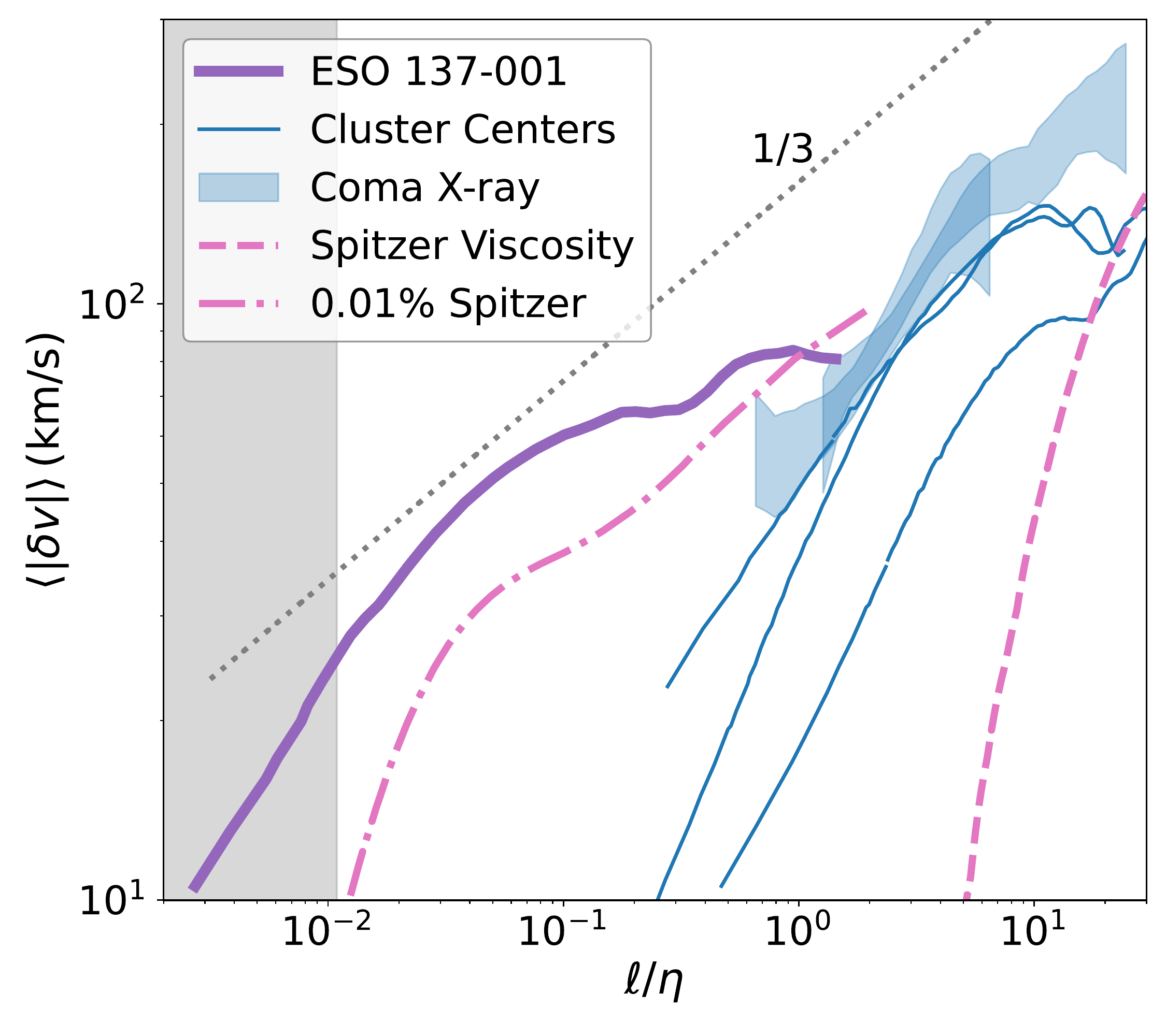}
\caption{VSF normalized by the Kolmogorov microscale $\eta$. The part at large $\ell$ with large sampling limits is removed for visual clarity. Also plotted are normalized VSFs of cluster center filaments (blue lines) and Chandra X-ray surface brightness fluctuation of Coma \citep{Zhuravleva2019}. All observations point to suppressed isotropic viscosity in the ICM. The turbulence we measure in ESO 137-001 suggests that ICM viscosity is below 0.01\% Spitzer.
}\label{fig:viscosity}
\end{figure}

Figure~\ref{fig:viscosity} shows the VSF of ESO 137-001 with $\ell$ normalized by the Kolmogorov microscale $\eta$. $\eta$ is computed as $\eta=\left (\frac{\nu^3}{\epsilon}\right)^{1/4}$, where $\nu$ is the kinematic viscosity and $\epsilon$ is the energy dissipation rate. The dynamic viscosity $\mu$, which is related to the kinetic viscosity as $\mu=\rho \nu$, can be estimated as:
\begin{equation}
\mu=5500\, \rm g\, cm^{-1} s^{-1} \left(\frac{T_{\rm e}}{10^8 K} \right)^{5/2} \left(\frac{ln \Lambda}{40}\right)^{-1} \,,
\end{equation}
where $\rm ln \Lambda$ is the Coulomb logarithm \citep{Sarazin1988}. For the ICM properties, we use $T_{\rm e}=6.3$ keV and $n_{\rm e}=1.3 \times10^{-3} \rm cm^{-3}$ \citep{Sun2010}. We use $\ell = 1$ kpc and $v_\ell=50$ km/s to obtain $\epsilon \sim 10^{-28} \rm erg\,s^{-1}\,cm^{-3}$. This gives us $\eta \sim 22$ kpc. We choose a scale $\ell$ large enough such that the VSF is not affected by steepening due to smoothing (see Section~\ref{sec:uncertainties} for details). The VSFs with and without rotation are also converged at this scale, making the results independent of the rotation modeling. We have verified that the results are not sensitive to exactly where $\ell$ and $v_\ell$ are measured. Even if we use the VSF of the molecular gas with $\ell = 1$ kpc and $v_\ell=40$ km/s, the computed $\eta$ only changes by less than 20\%.

For reference, we also plot the expectations based on direct numerical simulations \citep{Ishihara2016} for Spitzer viscosity and 0.01\% Spitzer in Figure~\ref{fig:viscosity}. Details of how the pink lines in Figure~\ref{fig:viscosity} are computed can be found in \citet{Zhuravleva2019}. The entire dynamical range we probe here is below the dissipation scale if ICM viscosity is Spitzer. 
Previous analyses using H$\alpha$ filaments in cluster centers were able to constrain ICM viscosity to $\sim 1$\% Spitzer level. Previous best constraints based on the X-ray data of Coma show a suppression of $0.1-10\%$ depending on the Prandtl number \citep{Zhuravleva2019}.
The measured turbulence in ESO 137-001 is more consistent with at least a suppression of $\sim 0.01$\% Spitzer level. 
Our result suggests that the macroscopic properties of the ICM is strongly modified by microscopic plasma processes operating much below the mean free path.

\section{Conclusions and Final Remarks}\label{sec:conclusions}

We use high-resolution MUSE and ALMA observations to study the kinematics of the multiphase gas in ESO 137-001, a quintessential jellyfish galaxy. We compute the first-order VSF and show that the motion of the multiphase gas is consistent with the expectation of a turbulent flow.
Along the tail direction, the level of turbulence builds up from ISM turbulence in the ``head'' to stronger turbulence driven by KH instability that saturates at $\sim 30$ kpc. There is also a hint of inherited large-scale ICM turbulence toward the end of the tail. The ionized gas observed with MUSE and the molecular gas observed with ALMA appear reasonably well-coupled kinematically, although the level of turbulence is slightly lower in the latter phase. Using cool gas as kinematic tracers of the hot ICM, we find that isotropic viscosity has to be suppressed to below $0.01\%$ Spitzer level. Future numerical studies can help better understand the importance of different drivers of turbulence, the coupling between different phases, and the role of ICM viscosity in the kinematics of jellyfish tails.

\section*{Acknowledgments}
We would like to thank Irina Zhuravleva, Thomas Berlok, Stephanie Tonnesen, Greg Bryan, Eliot Quataert, Yuanyuan Su, Valeria Olivares, and Trung Ha for helpful discussions. 
Y.L. acknowledges financial support from NSF grants AST-2107735 and AST-2219686, NASA grant 80NSSC22K0668, and Chandra X-ray Observatory grant TM3-24005X. 
M.S. acknowledges support from the NSF grant 1714764 and the NASA grants 80NSSC21K0704 and 80NSSC19K1257. 
P.J. acknowledges support from the project RVO:67985815, and the project LM2023059 of the Ministry of Education, Youth and Sports of the Czech Republic.
This work was partly performed at the Aspen Center for Physics, which is supported by National Science Foundation grant PHY-1607611. 

\section*{Data Availability}

The data underlying this article will be shared on reasonable request to the corresponding author.

\bibliographystyle{mnras}
%\bibliography{library}

\begin{thebibliography}{}
\makeatletter
\relax
\def\mn@urlcharsother{\let\do\@makeother \do\$\do\&\do\#\do\^\do\_\do\%\do\~}
\def\mn@doi{\begingroup\mn@urlcharsother \@ifnextchar [ {\mn@doi@}
  {\mn@doi@[]}}
\def\mn@doi@[#1]#2{\def\@tempa{#1}\ifx\@tempa\@empty \href
  {http://dx.doi.org/#2} {doi:#2}\else \href {http://dx.doi.org/#2} {#1}\fi
  \endgroup}
\def\mn@eprint#1#2{\mn@eprint@#1:#2::\@nil}
\def\mn@eprint@arXiv#1{\href {http://arxiv.org/abs/#1} {{\tt arXiv:#1}}}
\def\mn@eprint@dblp#1{\href {http://dblp.uni-trier.de/rec/bibtex/#1.xml}
  {dblp:#1}}
\def\mn@eprint@#1:#2:#3:#4\@nil{\def\@tempa {#1}\def\@tempb {#2}\def\@tempc
  {#3}\ifx \@tempc \@empty \let \@tempc \@tempb \let \@tempb \@tempa \fi \ifx
  \@tempb \@empty \def\@tempb {arXiv}\fi \@ifundefined
  {mn@eprint@\@tempb}{\@tempb:\@tempc}{\expandafter \expandafter \csname
  mn@eprint@\@tempb\endcsname \expandafter{\@tempc}}}

\bibitem[\protect\citeauthoryear{{Abruzzo}, {Bryan}  \& {Fielding}}{{Abruzzo}
  et~al.}{2022}]{2022ApJ...925..199A}
{Abruzzo} M.~W.,  {Bryan} G.~L.,   {Fielding} D.~B.,  2022, \mn@doi [\apj]
  {10.3847/1538-4357/ac3c48}, \href
  {https://ui.adsabs.harvard.edu/abs/2022ApJ...925..199A} {925, 199}

\bibitem[\protect\citeauthoryear{{Angelinelli}, {Vazza}, {Giocoli}, {Ettori},
  {Jones}, {Brunetti}, {Br{\"u}ggen}  \& {Eckert}}{{Angelinelli}
  et~al.}{2020}]{Angelinelli2020}
{Angelinelli} M.,  {Vazza} F.,  {Giocoli} C.,  {Ettori} S.,  {Jones} T.~W.,
  {Brunetti} G.,  {Br{\"u}ggen} M.,   {Eckert} D.,  2020, \mn@doi [\mnras]
  {10.1093/mnras/staa975}, \href
  {https://ui.adsabs.harvard.edu/abs/2020MNRAS.495..864A} {495, 864}

\bibitem[\protect\citeauthoryear{{Arzamasskiy}, {Kunz}, {Squire}, {Quataert}
  \& {Schekochihin}}{{Arzamasskiy} et~al.}{2022}]{2022arXiv220705189A}
{Arzamasskiy} L.,  {Kunz} M.~W.,  {Squire} J.,  {Quataert} E.,   {Schekochihin}
  A.~A.,  2022, \mn@doi [arXiv e-prints] {10.48550/arXiv.2207.05189}, \href
  {https://ui.adsabs.harvard.edu/abs/2022arXiv220705189A} {p. arXiv:2207.05189}

\bibitem[\protect\citeauthoryear{{Berlok} \& {Pfrommer}}{{Berlok} \&
  {Pfrommer}}{2019}]{Berlok2019}
{Berlok} T.,  {Pfrommer} C.,  2019, \mn@doi [\mnras] {10.1093/mnras/stz2347},
  \href {https://ui.adsabs.harvard.edu/abs/2019MNRAS.489.3368B} {489, 3368}

\bibitem[\protect\citeauthoryear{{Boselli}, {Fossati}  \& {Sun}}{{Boselli}
  et~al.}{2022}]{2022A&ARv..30....3B}
{Boselli} A.,  {Fossati} M.,   {Sun} M.,  2022, \mn@doi [\aapr]
  {10.1007/s00159-022-00140-3}, \href
  {https://ui.adsabs.harvard.edu/abs/2022A&ARv..30....3B} {30, 3}

\bibitem[\protect\citeauthoryear{{Chen} et~al.,}{{Chen}
  et~al.}{2022}]{Chen2022}
{Chen} M.~C.,  et~al., 2022, arXiv e-prints, \href
  {https://ui.adsabs.harvard.edu/abs/2022arXiv220904344C} {p. arXiv:2209.04344}

\bibitem[\protect\citeauthoryear{{Cottle}, {Scannapieco}, {Br{\"u}ggen},
  {Banda-Barrag{\'a}n}  \& {Federrath}}{{Cottle}
  et~al.}{2020}]{2020ApJ...892...59C}
{Cottle} J.,  {Scannapieco} E.,  {Br{\"u}ggen} M.,  {Banda-Barrag{\'a}n} W.,
  {Federrath} C.,  2020, \mn@doi [\apj] {10.3847/1538-4357/ab76d1}, \href
  {https://ui.adsabs.harvard.edu/abs/2020ApJ...892...59C} {892, 59}

\bibitem[\protect\citeauthoryear{{Dolag}, {Vazza}, {Brunetti}  \&
  {Tormen}}{{Dolag} et~al.}{2005}]{Dolag2005}
{Dolag} K.,  {Vazza} F.,  {Brunetti} G.,   {Tormen} G.,  2005, \mn@doi [\mnras]
  {10.1111/j.1365-2966.2005.09630.x}, \href
  {https://ui.adsabs.harvard.edu/abs/2005MNRAS.364..753D} {364, 753}

\bibitem[\protect\citeauthoryear{{Farber}, {Ruszkowski}, {Tonnesen}  \&
  {Holguin}}{{Farber} et~al.}{2022}]{2022MNRAS.512.5927F}
{Farber} R.~J.,  {Ruszkowski} M.,  {Tonnesen} S.,   {Holguin} F.,  2022,
  \mn@doi [\mnras] {10.1093/mnras/stac794}, \href
  {https://ui.adsabs.harvard.edu/abs/2022MNRAS.512.5927F} {512, 5927}

\bibitem[\protect\citeauthoryear{{Fossati}, {Fumagalli}, {Boselli}, {Gavazzi},
  {Sun}  \& {Wilman}}{{Fossati} et~al.}{2016}]{Fossati2016}
{Fossati} M.,  {Fumagalli} M.,  {Boselli} A.,  {Gavazzi} G.,  {Sun} M.,
  {Wilman} D.~J.,  2016, \mn@doi [\mnras] {10.1093/mnras/stv2400}, \href
  {https://ui.adsabs.harvard.edu/abs/2016MNRAS.455.2028F} {455, 2028}

\bibitem[\protect\citeauthoryear{{Fumagalli}, {Fossati}, {Hau}, {Gavazzi},
  {Bower}, {Sun}  \& {Boselli}}{{Fumagalli} et~al.}{2014}]{Fossati2014}
{Fumagalli} M.,  {Fossati} M.,  {Hau} G. K.~T.,  {Gavazzi} G.,  {Bower} R.,
  {Sun} M.,   {Boselli} A.,  2014, \mn@doi [\mnras] {10.1093/mnras/stu2092},
  \href {https://ui.adsabs.harvard.edu/abs/2014MNRAS.445.4335F} {445, 4335}

\bibitem[\protect\citeauthoryear{{Ganguly} et~al.,}{{Ganguly}
  et~al.}{2023}]{Ganguly2023}
{Ganguly} S.,  et~al., 2023, submitted

\bibitem[\protect\citeauthoryear{{Ge} et~al.,}{{Ge} et~al.}{2021}]{Ge2021}
{Ge} C.,  et~al., 2021, \mn@doi [\mnras] {10.1093/mnras/stab1569}, \href
  {https://ui.adsabs.harvard.edu/abs/2021MNRAS.505.4702G} {505, 4702}

\bibitem[\protect\citeauthoryear{{Gendron-Marsolais}
  et~al.,}{{Gendron-Marsolais} et~al.}{2018}]{Gendron-Marsolais2018}
{Gendron-Marsolais} M.,  et~al., 2018, \mn@doi [\mnras]
  {10.1093/mnrasl/sly084}, \href
  {https://ui.adsabs.harvard.edu/abs/2018MNRAS.479L..28G} {479, L28}

\bibitem[\protect\citeauthoryear{{Gronke} \& {Oh}}{{Gronke} \&
  {Oh}}{2018}]{2018MNRAS.480L.111G}
{Gronke} M.,  {Oh} S.~P.,  2018, \mn@doi [\mnras] {10.1093/mnrasl/sly131},
  \href {https://ui.adsabs.harvard.edu/abs/2018MNRAS.480L.111G} {480, L111}

\bibitem[\protect\citeauthoryear{{Gronke}, {Oh}, {Ji}  \& {Norman}}{{Gronke}
  et~al.}{2022}]{Gronke2022}
{Gronke} M.,  {Oh} S.~P.,  {Ji} S.,   {Norman} C.,  2022, \mn@doi [\mnras]
  {10.1093/mnras/stab3351}, \href
  {https://ui.adsabs.harvard.edu/abs/2022MNRAS.511..859G} {511, 859}

\bibitem[\protect\citeauthoryear{{Ha}, {Li}, {Xu}, {Kounkel}  \& {Li}}{{Ha}
  et~al.}{2021}]{Ha2021}
{Ha} T.,  {Li} Y.,  {Xu} S.,  {Kounkel} M.,   {Li} H.,  2021, \mn@doi [\apjl]
  {10.3847/2041-8213/abd8c9}, \href
  {https://ui.adsabs.harvard.edu/abs/2021ApJ...907L..40H} {907, L40}

\bibitem[\protect\citeauthoryear{{Ha}, {Li}, {Kounkel}, {Xu}, {Li}  \&
  {Zheng}}{{Ha} et~al.}{2022}]{Ha2022}
{Ha} T.,  {Li} Y.,  {Kounkel} M.,  {Xu} S.,  {Li} H.,   {Zheng} Y.,  2022,
  \mn@doi [\apj] {10.3847/1538-4357/ac76bf}, \href
  {https://ui.adsabs.harvard.edu/abs/2022ApJ...934....7H} {934, 7}

\bibitem[\protect\citeauthoryear{{Ishihara}, {Morishita}, {Yokokawa}, {Uno}  \&
  {Kaneda}}{{Ishihara} et~al.}{2016}]{Ishihara2016}
{Ishihara} T.,  {Morishita} K.,  {Yokokawa} M.,  {Uno} A.,   {Kaneda} Y.,
  2016, \mn@doi [Physical Review Fluids] {10.1103/PhysRevFluids.1.082403},
  \href {https://ui.adsabs.harvard.edu/abs/2016PhRvF...1h2403I} {1, 082403}

\bibitem[\protect\citeauthoryear{{J{\'a}chym}, {Combes}, {Cortese}, {Sun}  \&
  {Kenney}}{{J{\'a}chym} et~al.}{2014}]{Pavel2014}
{J{\'a}chym} P.,  {Combes} F.,  {Cortese} L.,  {Sun} M.,   {Kenney} J. D.~P.,
  2014, \mn@doi [\apj] {10.1088/0004-637X/792/1/11}, \href
  {https://ui.adsabs.harvard.edu/abs/2014ApJ...792...11J} {792, 11}

\bibitem[\protect\citeauthoryear{{J{\'a}chym} et~al.,}{{J{\'a}chym}
  et~al.}{2019}]{Pavel2019}
{J{\'a}chym} P.,  et~al., 2019, \mn@doi [\apj] {10.3847/1538-4357/ab3e6c},
  \href {https://ui.adsabs.harvard.edu/abs/2019ApJ...883..145J} {883, 145}

\bibitem[\protect\citeauthoryear{{Ji}, {Oh}  \& {Masterson}}{{Ji}
  et~al.}{2019}]{2019MNRAS.487..737J}
{Ji} S.,  {Oh} S.~P.,   {Masterson} P.,  2019, \mn@doi [\mnras]
  {10.1093/mnras/stz1248}, \href
  {https://ui.adsabs.harvard.edu/abs/2019MNRAS.487..737J} {487, 737}

\bibitem[\protect\citeauthoryear{{Kunz}, {Schekochihin}  \& {Stone}}{{Kunz}
  et~al.}{2014}]{Kunz2014}
{Kunz} M.~W.,  {Schekochihin} A.~A.,   {Stone} J.~M.,  2014, \mn@doi [Physical
  Review Letters] {10.1103/PhysRevLett.112.205003}, \href
  {https://ui.adsabs.harvard.edu/abs/2014PhRvL.112t5003K} {112, 205003}

\bibitem[\protect\citeauthoryear{{Larson}}{{Larson}}{1981}]{Larson1981}
{Larson} R.~B.,  1981, \mn@doi [\mnras] {10.1093/mnras/194.4.809}, \href
  {https://ui.adsabs.harvard.edu/abs/1981MNRAS.194..809L} {194, 809}

\bibitem[\protect\citeauthoryear{{Li}, {Hopkins}, {Squire}  \& {Hummels}}{{Li}
  et~al.}{2020a}]{2020MNRAS.492.1841L}
{Li} Z.,  {Hopkins} P.~F.,  {Squire} J.,   {Hummels} C.,  2020a, \mn@doi
  [\mnras] {10.1093/mnras/stz3567}, \href
  {https://ui.adsabs.harvard.edu/abs/2020MNRAS.492.1841L} {492, 1841}

\bibitem[\protect\citeauthoryear{{Li} et~al.,}{{Li} et~al.}{2020b}]{Li2020}
{Li} Y.,  et~al., 2020b, \mn@doi [\apjl] {10.3847/2041-8213/ab65c7}, \href
  {https://ui.adsabs.harvard.edu/abs/2020ApJ...889L...1L} {889, L1}

\bibitem[\protect\citeauthoryear{{Luo} et~al.,}{{Luo} et~al.}{2022}]{Luo2022}
{Luo} R.,  et~al., 2022, \mn@doi [arXiv e-prints] {10.48550/arXiv.2212.03891},
  \href {https://ui.adsabs.harvard.edu/abs/2022arXiv221203891L} {p.
  arXiv:2212.03891}

\bibitem[\protect\citeauthoryear{{Mandelker}, {Nagai}, {Aung}, {Dekel},
  {Birnboim}  \& {van den Bosch}}{{Mandelker} et~al.}{2020}]{Mandelker2020}
{Mandelker} N.,  {Nagai} D.,  {Aung} H.,  {Dekel} A.,  {Birnboim} Y.,   {van
  den Bosch} F.~C.,  2020, \mn@doi [\mnras] {10.1093/mnras/staa812}, \href
  {https://ui.adsabs.harvard.edu/abs/2020MNRAS.494.2641M} {494, 2641}

\bibitem[\protect\citeauthoryear{{Mohapatra}, {Jetti}, {Sharma}  \&
  {Federrath}}{{Mohapatra} et~al.}{2022}]{Mohapatra2022}
{Mohapatra} R.,  {Jetti} M.,  {Sharma} P.,   {Federrath} C.,  2022, \mn@doi
  [\mnras] {10.1093/mnras/stab3429}, \href
  {https://ui.adsabs.harvard.edu/abs/2022MNRAS.510.2327M} {510, 2327}

\bibitem[\protect\citeauthoryear{{Qian}, {Li}, {Offner}  \& {Pan}}{{Qian}
  et~al.}{2015}]{Qian2015}
{Qian} L.,  {Li} D.,  {Offner} S.,   {Pan} Z.,  2015, \mn@doi [\apj]
  {10.1088/0004-637X/811/1/71}, \href
  {https://ui.adsabs.harvard.edu/abs/2015ApJ...811...71Q} {811, 71}

\bibitem[\protect\citeauthoryear{{Roediger} \& {Br{\"u}ggen}}{{Roediger} \&
  {Br{\"u}ggen}}{2006}]{Roediger2006}
{Roediger} E.,  {Br{\"u}ggen} M.,  2006, \mn@doi [\mnras]
  {10.1111/j.1365-2966.2006.10335.x}, \href
  {https://ui.adsabs.harvard.edu/abs/2006MNRAS.369..567R} {369, 567}

\bibitem[\protect\citeauthoryear{{Roediger} et~al.,}{{Roediger}
  et~al.}{2015}]{2015ApJ...806..104R}
{Roediger} E.,  et~al., 2015, \mn@doi [\apj] {10.1088/0004-637X/806/1/104},
  \href {https://ui.adsabs.harvard.edu/abs/2015ApJ...806..104R} {806, 104}

\bibitem[\protect\citeauthoryear{{Ruszkowski}, {Br{\"u}ggen}, {Lee}  \&
  {Shin}}{{Ruszkowski} et~al.}{2014}]{Ruszkowski2014}
{Ruszkowski} M.,  {Br{\"u}ggen} M.,  {Lee} D.,   {Shin} M.~S.,  2014, \mn@doi
  [\apj] {10.1088/0004-637X/784/1/75}, \href
  {https://ui.adsabs.harvard.edu/abs/2014ApJ...784...75R} {784, 75}

\bibitem[\protect\citeauthoryear{{Ryu}, {Kang}, {Cho}  \& {Das}}{{Ryu}
  et~al.}{2008}]{Ryu2008}
{Ryu} D.,  {Kang} H.,  {Cho} J.,   {Das} S.,  2008, \mn@doi [Science]
  {10.1126/science.1154923}, \href
  {https://ui.adsabs.harvard.edu/abs/2008Sci...320..909R} {320, 909}

\bibitem[\protect\citeauthoryear{{Sarazin}}{{Sarazin}}{1988}]{Sarazin1988}
{Sarazin} C.~L.,  1988, {X-ray emission from clusters of galaxies}

\bibitem[\protect\citeauthoryear{{Shi}, {Nagai}  \& {Lau}}{{Shi}
  et~al.}{2018}]{Shi2018}
{Shi} X.,  {Nagai} D.,   {Lau} E.~T.,  2018, \mn@doi [\mnras]
  {10.1093/mnras/sty2340}, \href
  {https://ui.adsabs.harvard.edu/abs/2018MNRAS.481.1075S} {481, 1075}

\bibitem[\protect\citeauthoryear{{Sivanandam}, {Rieke}  \&
  {Rieke}}{{Sivanandam} et~al.}{2010}]{2010ApJ...717..147S}
{Sivanandam} S.,  {Rieke} M.~J.,   {Rieke} G.~H.,  2010, \mn@doi [\apj]
  {10.1088/0004-637X/717/1/147}, \href
  {https://ui.adsabs.harvard.edu/abs/2010ApJ...717..147S} {717, 147}

\bibitem[\protect\citeauthoryear{{Su} et~al.,}{{Su}
  et~al.}{2017}]{2017ApJ...834...74S}
{Su} Y.,  et~al., 2017, \mn@doi [\apj] {10.3847/1538-4357/834/1/74}, \href
  {https://ui.adsabs.harvard.edu/abs/2017ApJ...834...74S} {834, 74}

\bibitem[\protect\citeauthoryear{{Sun}, {Jones}, {Forman}, {Nulsen}, {Donahue}
  \& {Voit}}{{Sun} et~al.}{2006}]{Sun2006}
{Sun} M.,  {Jones} C.,  {Forman} W.,  {Nulsen} P.~E.~J.,  {Donahue} M.,
  {Voit} G.~M.,  2006, \mn@doi [\apjl] {10.1086/500590}, \href
  {https://ui.adsabs.harvard.edu/abs/2006ApJ...637L..81S} {637, L81}

\bibitem[\protect\citeauthoryear{{Sun}, {Donahue}  \& {Voit}}{{Sun}
  et~al.}{2007}]{Sun2007}
{Sun} M.,  {Donahue} M.,   {Voit} G.~M.,  2007, \mn@doi [\apj]
  {10.1086/522690}, \href
  {https://ui.adsabs.harvard.edu/abs/2007ApJ...671..190S} {671, 190}

\bibitem[\protect\citeauthoryear{{Sun}, {Donahue}, {Roediger}, {Nulsen},
  {Voit}, {Sarazin}, {Forman}  \& {Jones}}{{Sun} et~al.}{2010}]{Sun2010}
{Sun} M.,  {Donahue} M.,  {Roediger} E.,  {Nulsen} P.~E.~J.,  {Voit} G.~M.,
  {Sarazin} C.,  {Forman} W.,   {Jones} C.,  2010, \mn@doi [\apj]
  {10.1088/0004-637X/708/2/946}, \href
  {https://ui.adsabs.harvard.edu/abs/2010ApJ...708..946S} {708, 946}

\bibitem[\protect\citeauthoryear{{Tonnesen} \& {Bryan}}{{Tonnesen} \&
  {Bryan}}{2012}]{Tonnesen2012}
{Tonnesen} S.,  {Bryan} G.~L.,  2012, \mn@doi [\mnras]
  {10.1111/j.1365-2966.2012.20737.x}, \href
  {https://ui.adsabs.harvard.edu/abs/2012MNRAS.422.1609T} {422, 1609}

\bibitem[\protect\citeauthoryear{{Tonnesen} \& {Stone}}{{Tonnesen} \&
  {Stone}}{2014}]{Tonnesen2014}
{Tonnesen} S.,  {Stone} J.,  2014, \mn@doi [\apj]
  {10.1088/0004-637X/795/2/148}, \href
  {https://ui.adsabs.harvard.edu/abs/2014ApJ...795..148T} {795, 148}

\bibitem[\protect\citeauthoryear{{Troncoso-Iribarren}, {Padilla}, {Santander},
  {Lagos}, {Garc{\'\i}a-Lambas}, {Rodr{\'\i}guez}  \&
  {Contreras}}{{Troncoso-Iribarren} et~al.}{2020}]{Troncoso-Iribarren2020}
{Troncoso-Iribarren} P.,  {Padilla} N.,  {Santander} C.,  {Lagos} C.~D.~P.,
  {Garc{\'\i}a-Lambas} D.,  {Rodr{\'\i}guez} S.,   {Contreras} S.,  2020,
  \mn@doi [\mnras] {10.1093/mnras/staa274}, \href
  {https://ui.adsabs.harvard.edu/abs/2020MNRAS.497.4145T} {497, 4145}

\bibitem[\protect\citeauthoryear{{Vazza}, {Brunetti}, {Kritsuk}, {Wagner},
  {Gheller}  \& {Norman}}{{Vazza} et~al.}{2009}]{Vazza2009}
{Vazza} F.,  {Brunetti} G.,  {Kritsuk} A.,  {Wagner} R.,  {Gheller} C.,
  {Norman} M.,  2009, \mn@doi [\aap] {10.1051/0004-6361/200912535}, \href
  {https://ui.adsabs.harvard.edu/abs/2009A&A...504...33V} {504, 33}

\bibitem[\protect\citeauthoryear{{Wang}, {Ruszkowski}, {Pfrommer}, {Oh}  \&
  {Yang}}{{Wang} et~al.}{2021}]{Wang2021}
{Wang} C.,  {Ruszkowski} M.,  {Pfrommer} C.,  {Oh} S.~P.,   {Yang} H. Y.~K.,
  2021, \mn@doi [\mnras] {10.1093/mnras/stab966}, \href
  {https://ui.adsabs.harvard.edu/abs/2021MNRAS.504..898W} {504, 898}

\bibitem[\protect\citeauthoryear{{Xu}}{{Xu}}{2020}]{Xu2020}
{Xu} S.,  2020, \mn@doi [\mnras] {10.1093/mnras/stz3092}, \href
  {https://ui.adsabs.harvard.edu/abs/2020MNRAS.492.1044X} {492, 1044}

\bibitem[\protect\citeauthoryear{{Xu} \& {Zhang}}{{Xu} \&
  {Zhang}}{2020}]{Xu2020FRB}
{Xu} S.,  {Zhang} B.,  2020, \mn@doi [\apjl] {10.3847/2041-8213/aba760}, \href
  {https://ui.adsabs.harvard.edu/abs/2020ApJ...898L..48X} {898, L48}

\bibitem[\protect\citeauthoryear{{Yun} et~al.,}{{Yun} et~al.}{2019}]{Yun2019}
{Yun} K.,  et~al., 2019, \mn@doi [\mnras] {10.1093/mnras/sty3156}, \href
  {https://ui.adsabs.harvard.edu/abs/2019MNRAS.483.1042Y} {483, 1042}

\bibitem[\protect\citeauthoryear{{Zhang}, {Zhuravleva}, {Gendron-Marsolais},
  {Churazov}, {Schekochihin}  \& {Forman}}{{Zhang} et~al.}{2022}]{Zhang2022}
{Zhang} C.,  {Zhuravleva} I.,  {Gendron-Marsolais} M.-L.,  {Churazov} E.,
  {Schekochihin} A.~A.,   {Forman} W.~R.,  2022, \mn@doi [\mnras]
  {10.1093/mnras/stac2282}, \href
  {https://ui.adsabs.harvard.edu/abs/2022MNRAS.tmp.2178Z} {}

\bibitem[\protect\citeauthoryear{{Zhuravleva}, {Churazov}, {Schekochihin},
  {Allen}, {Vikhlinin}  \& {Werner}}{{Zhuravleva}
  et~al.}{2019}]{Zhuravleva2019}
{Zhuravleva} I.,  {Churazov} E.,  {Schekochihin} A.~A.,  {Allen} S.~W.,
  {Vikhlinin} A.,   {Werner} N.,  2019, \mn@doi [Nature Astronomy]
  {10.1038/s41550-019-0794-z}, \href
  {https://ui.adsabs.harvard.edu/abs/2019NatAs...3..832Z} {3, 832}

\bibitem[\protect\citeauthoryear{{ZuHone}, {Markevitch}, {Brunetti}  \&
  {Giacintucci}}{{ZuHone} et~al.}{2013}]{ZuHone2013}
{ZuHone} J.~A.,  {Markevitch} M.,  {Brunetti} G.,   {Giacintucci} S.,  2013,
  \mn@doi [\apj] {10.1088/0004-637X/762/2/78}, \href
  {https://ui.adsabs.harvard.edu/abs/2013ApJ...762...78Z} {762, 78}

\bibitem[\protect\citeauthoryear{{ZuHone}, {Markevitch}  \&
  {Zhuravleva}}{{ZuHone} et~al.}{2016}]{ZuHone2016}
{ZuHone} J.~A.,  {Markevitch} M.,   {Zhuravleva} I.,  2016, \mn@doi [\apj]
  {10.3847/0004-637X/817/2/110}, \href
  {https://ui.adsabs.harvard.edu/abs/2016ApJ...817..110Z} {817, 110}

\makeatother
\end{thebibliography}

\appendix

\section{The Impact of Velocity Uncertainties on the Measured VSFs}\label{sec:appendix}

\begin{figure}
\centering
\includegraphics[scale=0.35,trim=0cm 0cm 0cm 0cm, clip=true]{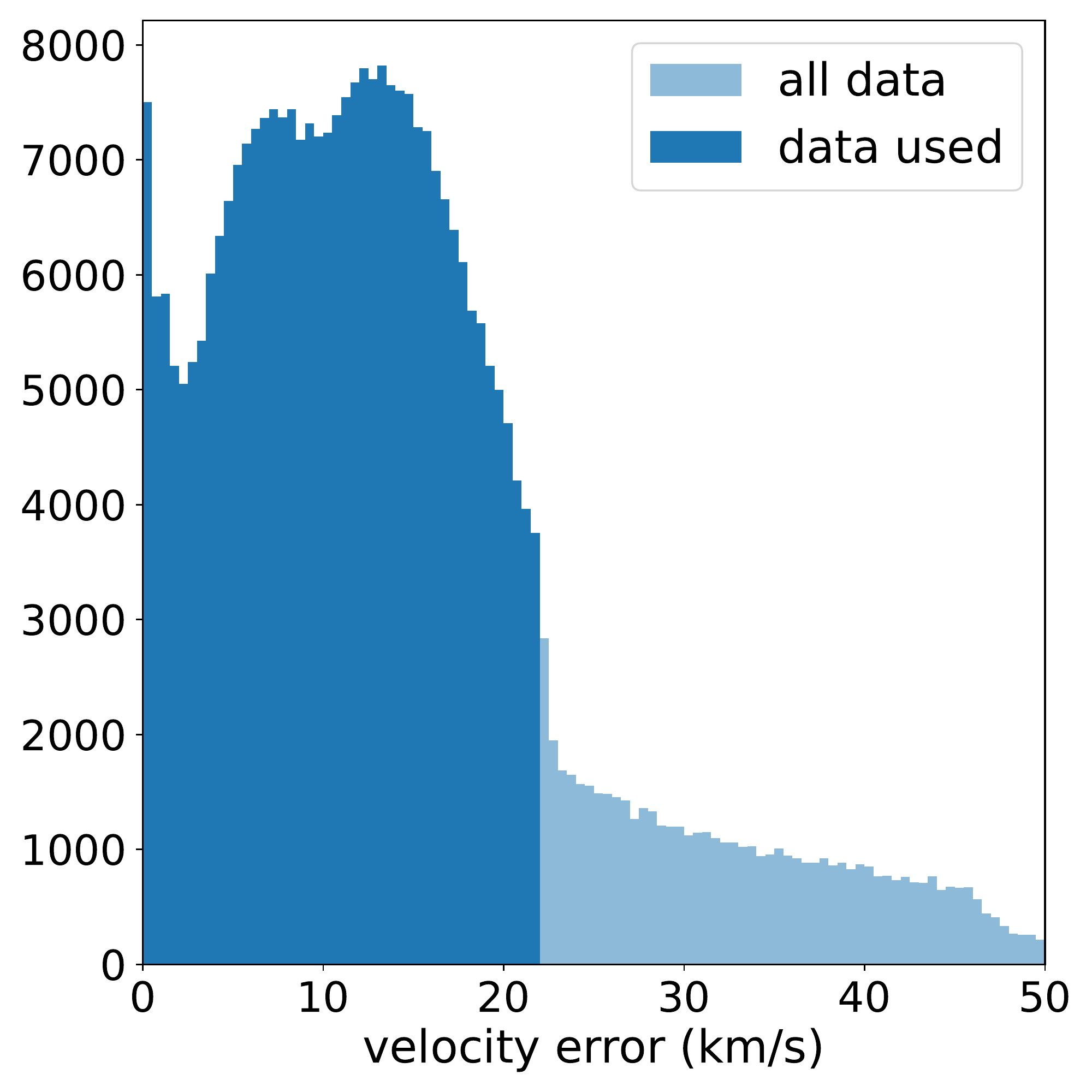}
\caption{The distribution of uncertainties in the measured line-of-sight velocities for all the MUSE H$\alpha$ data with SNR$>3$. 
}\label{fig:velo_error}
\end{figure}

We discuss how the measurement uncertainties in the line-of-sight velocities affect the VSFs here. 

Figure~\ref{fig:velo_error} shows the distribution of velocity errors in the measured line-of-sight velocities of the H$\alpha$ clouds with SNR$>3$. The SNR cut already removed many pixels with large velocity uncertainties. We then apply a velocity error cut at 22 km/s to further reduce the level of noise in our analysis. This is a natural choice based on the velocity error distribution and the value is similar to what is used in previous VSF studies using MUSE data on cluster center filaments \citep{Li2020}. Overall, the velocity uncertainties are small. 
However, velocity errors can still have an impact on the amplitude of the VSF on small scales. 

\begin{figure}
\centering
\includegraphics[scale=0.35,trim=0.5cm 0cm 0cm 0cm, clip=true]{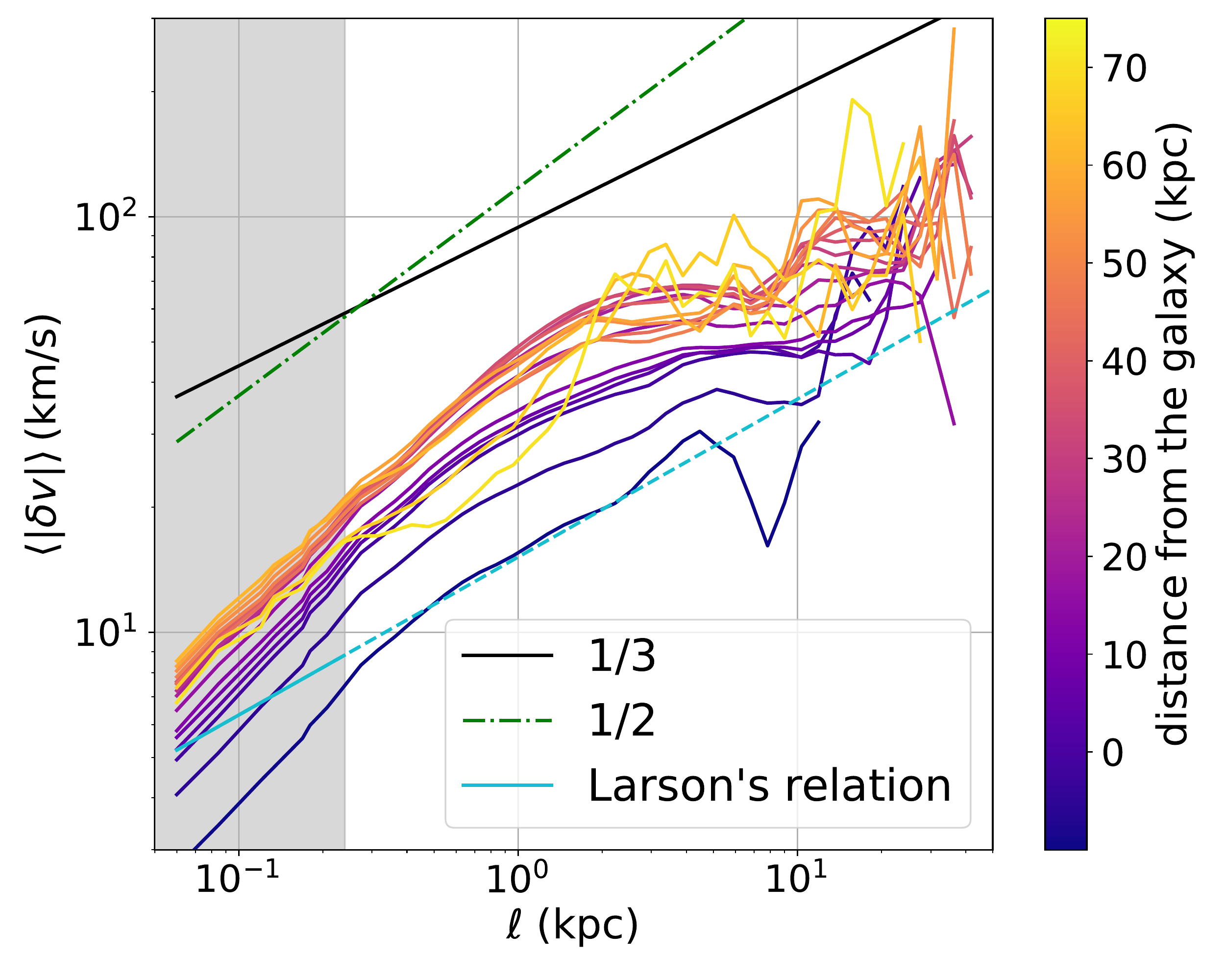}
\caption{The buildup of turbulence in the tail of ESO 137-001. Lines are made in the same way as Figure~\ref{fig:frames} but we apply a more strict velocity error cut of 10 km/s to the MUSE data instead of 22 km/s.}\label{fig:frames_error10}
\end{figure}

Figure~\ref{fig:frames_error10} shows the VSF analysis with the same moving frames but a more stringent velocity error cut at 10 km/s, effectively using only half of the data with high SNR. The main trends are the same as the original Figure~\ref{fig:frames}. 
The amplitude of the VSF on small scales has decreased when more noise is removed, but the VSF on large scales appears mostly unchanged. 
This is because the noise contributes to all scales with roughly equal power. The effect of noise thus becomes more negligible at larger scales with higher amplitudes in the signal.
More discussions on the effects of noise can be found in \citet{Ha2021, Ha2022} and \citet{Ganguly2023}.

% Don't change these lines
\bsp	% typesetting comment
\label{lastpage}
\end{document}